\documentclass[11pt]{article}
\usepackage{mor,epsfig}

\def\Journal#1#2#3#4{{#1} {\bf #2}, #3 (#4)}

\def\PLB{{\em Phys. Lett.}  B}

\def\PRD{{\em Phys. Rev.} D}
\def\ZPC{{\em Z. Phys.} C}

\def\be{\begin{equation}}
\def\ee{\end{equation}}
\def\bea{\begin{eqnarray}}
\def\eea{\end{eqnarray}}

\begin{document}
\vspace*{4cm}
\title{MEASUREMENTS OF DVCS AND HIGH ${\bf |\lowercase{t}|}$ VECTOR MESON PRODUCTION AT HERA}

\author{ D.P. BROWN \\ ~ \\ON BEHALF OF THE H1 AND ZEUS COLLABORATIONS\\}

\address{ ~\\ DESY, Notkestr. 85, D-22607 HAMBURG. GERMANY.}

\maketitle\abstracts{
Measurements of Deeply Virtual Compton Scattering (DVCS) are presented
using data collected by the H1 and ZEUS experiments at HERA.
The transition from a virtual photon to an on-shell particle is
expected to necessitate the use of skewed parton distributions for a
theoretical description.
The results are compared to theoretical predictions. Diffractive
vector meson production measurements from H1 and ZEUS are also
presented in which the magnitude of the four-momentum transfer
squared $|t|$ at the proton vertex is large. A large $|t|$ is expected
to make it possible to perform a perturbative calculation. The results
are compared to recent theoretical predictions performed within the
framework of perturbative QCD.}

\section{Introduction}
Measurements of the diffractive production of a photon 
and of the diffractive production of vector mesons cover a wide 
kinematic region at HERA. In perturbative QCD (pQCD)
diffractive interactions can be modelled in the proton rest
frame in which the incoming photon is assumed to fluctuate into a
quark-antiquark pair before interacting with the proton. The
interaction of the pair with the proton is via the exchange of a 
colour-singlet system. The detailed internal dynamical
structure of the system is studied in Deeply Virtual Compton 
Scattering (DVCS) and in diffractive meson production processes in which
the four-momentum transfer squared $|t|$ at the proton vertex is large.
DVCS measurements are expected to offer a particularly suitable channel for studying the 
behaviour of skewed parton distributions. The 
theoretical uncertainties associated with the meson
wavefunction which is not completely calculable in pQCD are also avoided.
Due to the experimentally clean signatures diffractive meson 
production processes at large $|t|$ have been proposed as ideal 
testing grounds for the dynamics of BFKL evolution.\cite{forshaw,bartels}

\section{Deeply Virtual Compton Scattering Measurements}

To enhance the ratio of selected DVCS
events to Bethe-Heitler events, which have the same experimental
signature ($ep \rightarrow e\gamma p$), the outgoing photon is selected in the 
forward, or outgoing proton, region.\cite{h1dvcs} Large values of
the incoming photon virtuality $Q^{2}$ are selected by detecting the scattered 
electron in the main detectors. The outgoing proton escapes down the 
beam-pipe in the forward direction. The remaining Bethe-Heitler contribution
is subtracted using a theoretical calculation.

\begin{figure}
%%\rule{5cm}{0.2mm}\hfill\rule{5cm}{0.2mm}
a) \hspace{7.0cm}b) \vskip -0.5cm
%%\rule{5cm}{0.2mm}\hfill\rule{5cm}{0.2mm}
\psfig{figure=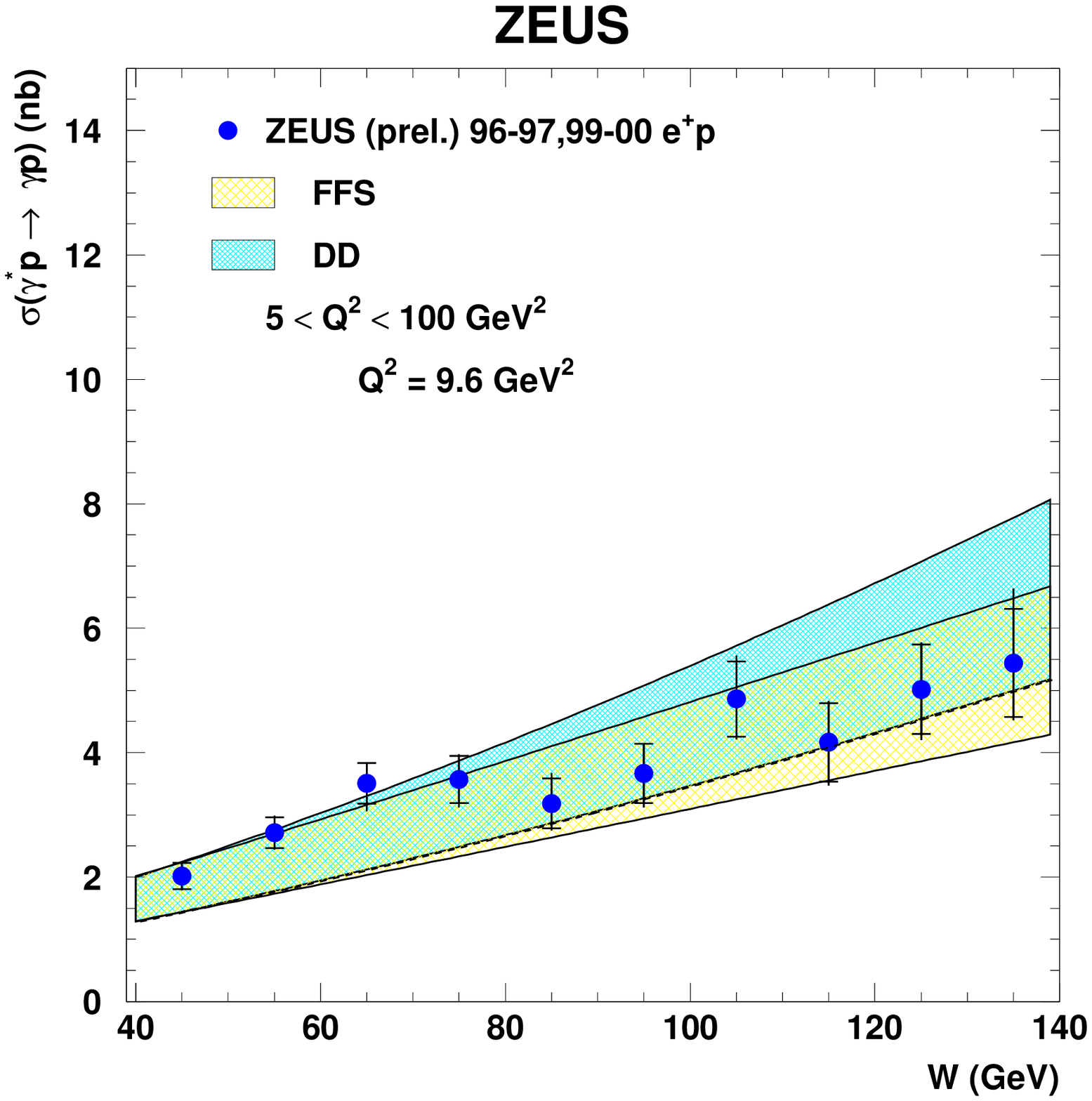,height=3.in}
\psfig{figure=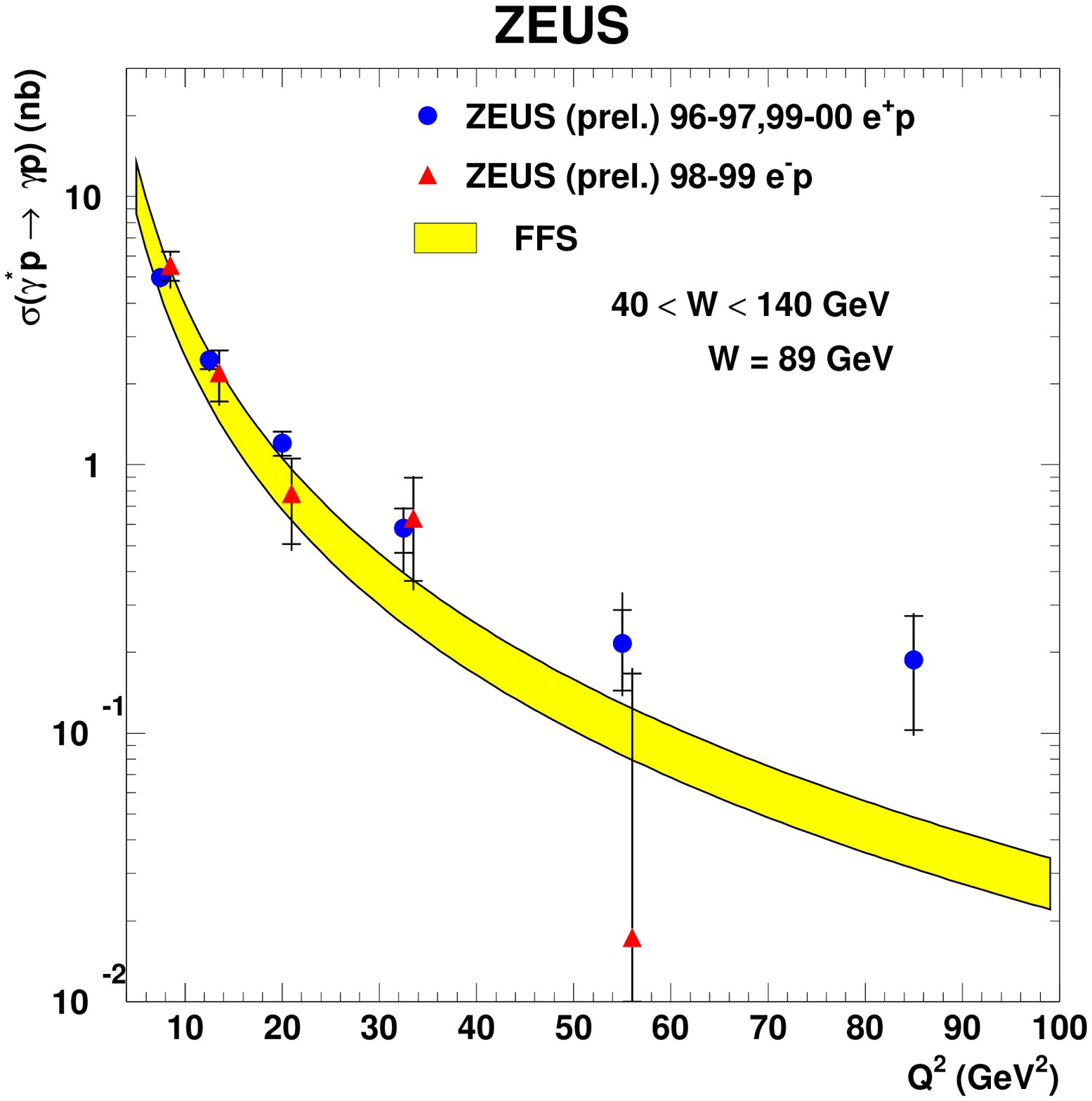,height=3.in}
\caption{The DVCS cross section $\sigma (\gamma^{*} p\rightarrow \gamma
  p)$ as a function of a) the photon-proton centre-of-mass energy
  $W_{\gamma p}$ and b) the incoming photon virtuality $Q^{2}$. The
  measurements are compared to DVCS models. The bands show the
  variations in the assumed $|t|$ dependence. \label{fig:dvcs}}
\end{figure}

The cross section \cite{zeusdvcs} for Deeply Virtual Compton Scattering $\sigma
(\gamma^{*} p \rightarrow \gamma p)$ extracted from positron-proton data 
is shown in figure \ref{fig:dvcs}a) as a function of the initial photon-proton
centre-of-mass energy $W_{\gamma p}$. The measurement
covers the kinematic region $40 < W_{\gamma p} < 140 ~\rm GeV$ and 
$5<Q^{2}<100 ~\rm GeV^{2}$.  A fit of the 
form $\sigma \sim W_{\gamma p}^{\delta}$ gives a similar steep rise $\delta =
0.78 \pm 0.10$ as measured for elastic $\rm J/\psi$
photoproduction.\cite{Adloff:2000vm,Chekanov:2002xi} The data are compared to theoretical predictions.\cite{ddosch,ffs}
Both models include soft and hard contributions and in both cases the 
bands show variations in the assumed $|t|$ dependence. Both models give a 
good description of the data. In figure \ref{fig:dvcs}b) the cross 
section is shown as a function of the incoming photon virtuality $Q^{2}$. 
The $e^{+}$ and $e^{-}$ analyses, both shown in the figure, are 
based on integrated luminosities of 95 $\rm pb^{-1}$ and 17 $\rm
pb^{-1}$ respectively. A fit of the form $(Q^{2})^{-n}$ yields an
exponent $n=1.47\pm 0.07$. The data are compared to the one of the models
\cite{ffs}, extrapolated under the assumption of a $Q^{2}$
independent $|t|$-dependence. The data lie above the prediction at large
values of $Q^{2}$.
 
\section{Diffractive Vector Meson Production Measurements at Large ${\bf |\lowercase{t}|}$}

\begin{figure}
a) \hspace{7.0cm}b) \vskip -0.5cm
\hspace{0.5cm}\psfig{figure=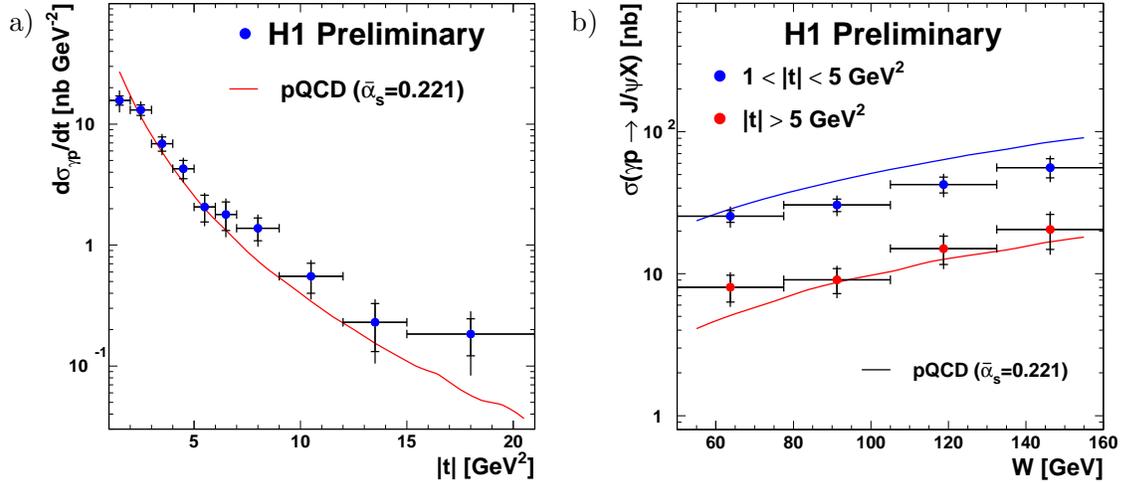,height=2.5in}
\hspace{1.0cm}\psfig{figure=figure2b.epsi,height=2.5in}
\caption{The photon-proton differential cross-section a)
${d\sigma/ d|t|}$ as a function of $|t|$ and  b) $\sigma (\gamma
p \rightarrow J/\psi X)$ as a function of $W_{\gamma p}$ for two
different intervals of ${|t|}$ for ${J/\psi}$ production with 
proton dissociation in the kinematic range
$Q^{2}<1.0~\mathrm{GeV^{2}}$, $50<W_{\gamma p}<160~\mathrm{GeV}$,
$|t|>1.0~\mathrm{GeV^{2}}$. The results are compared
to a BFKL model.
\label{fig:hvm}}
\vskip -0.3cm
\end{figure}

The mesons are selected via their decays 
into two-prong decay signatures.  The four-momentum
transfer squared $|t|$ at the proton vertex is reconstructed 
using the meson decay products. 
At large values of the $|t|$, the proton dissociates and the products 
typically leave energy deposits in the forward detectors.
The differential cross section $d\sigma/dt$ for the diffractive
photoproduction of the $J/\psi$ is shown as a function of $|t|$ 
extending to very large $|t|<21 ~\rm GeV^{2}$ in figure \ref{fig:hvm}a).\cite{large_t}
\newpage

In figure \ref{fig:hvm}b) the total cross 
section $\sigma(\gamma p \rightarrow J/\psi Y)$ is also presented as a 
function of $W_{\gamma p}$ in two $|t|$ regions. 
Fits of the form $\sigma \sim W_{\gamma  p}^{\delta}$ yield a 
similar steep rise ($\delta \sim 1$) as for elastic $J/\psi$
photoproduction at smaller values of
$|t|$.\cite{Adloff:2000vm,Chekanov:2002xi} The data are compared to a
pQCD prediction which uses BFKL
evolution and a simple non-relativistic treatment of the meson 
wavefunction.\cite{forshaw,bartels} The model gives a reasonable simultaneous 
description of the $W_{\gamma p}$ and $|t|$ distributions,
when the parameter $\bar{\alpha}_s$, treated as a free parameter 
in this model, is chosen to be about $\bar{\alpha}_s = 0.2$. 

\begin{figure}[b]
\vspace{7.5cm}
\hspace{2.5cm}\begin{picture}(14,6)
\put(5.0,-18.3){\epsfig{figure=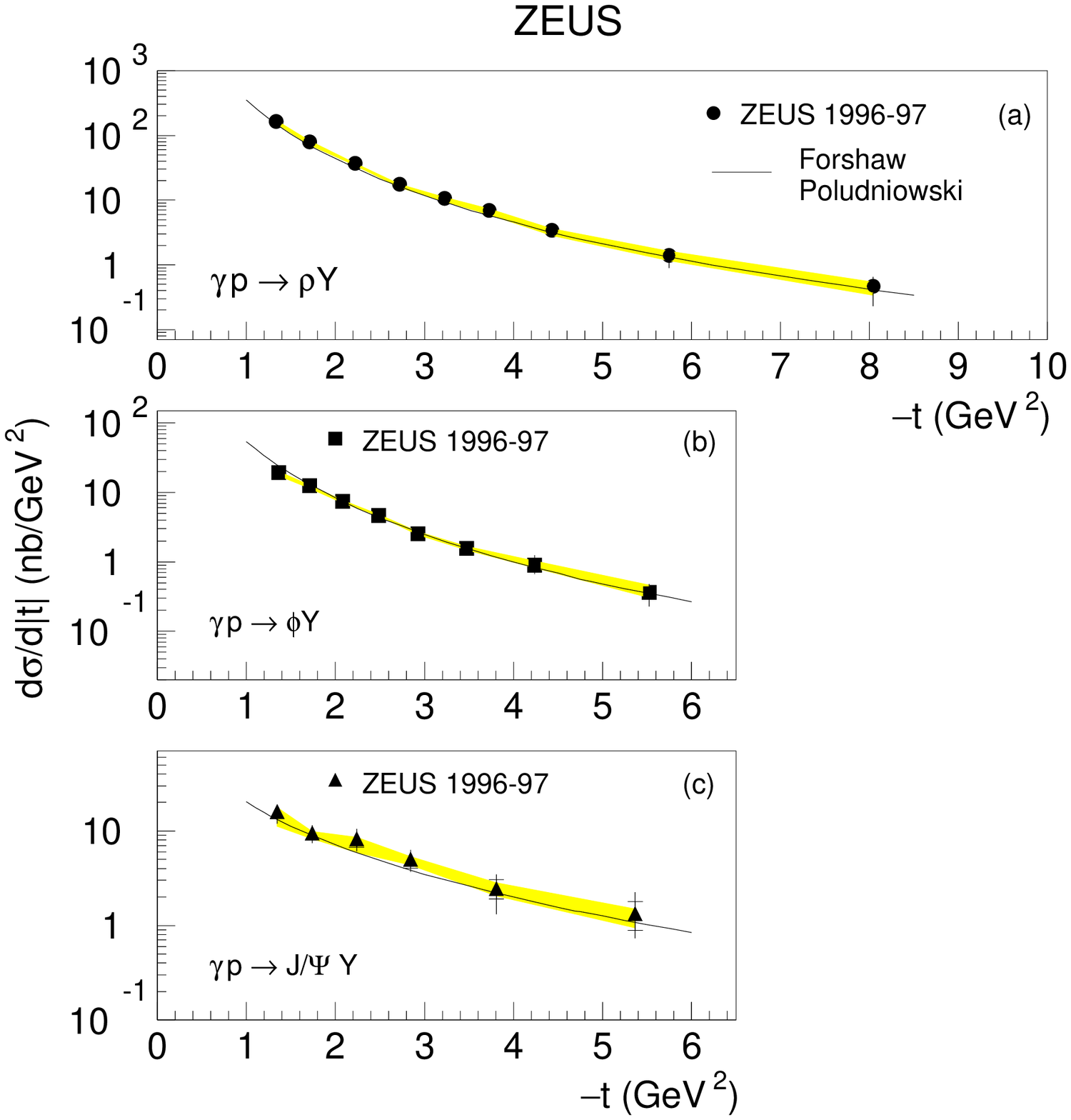,width=10.3cm}}
\put(5,-18.3){\epsfig{figure=whitebox.eps,width=11.3cm,height=3.5cm}}
\end{picture}
\vspace{-3.0cm}
\caption{The differential cross section $d\sigma /dt$ as a function of
  $|t|$ for the proton dissociative production of a) the $\rho$ and b)
  the $\phi$ meson. The result of a simultaneous fit of a BFKL model
  is shown by the solid line. \label{fig:rhophit}}
\end{figure}

The measured differential cross section $d\sigma/dt$ for the proton
dissociative production of the $\rho$ and the $\phi$ meson are 
shown in figure \ref{fig:rhophit}a) and b) respectively for the 
kinematic region $80 < W_{\gamma p} < 120 ~\rm GeV$, 
$Q^{2}<0.02 ~\rm GeV^{2}$, based on an integrated luminosity of
25 $\rm pb^{-1}$. The dependence at large $|t|$ is approximately
power-like $|t|^{-n}$ with an exponent $n \sim 3$ for both the $\rho$
and the $\phi$.\cite{Chekanov:2002rm} The $|t|$ dependence of the
$\rho$ and the $\phi$ are fitted simultaneously
\cite{Forshaw:2001pf} together with $J/\psi$ data at low values of
$|t|$ to the BFKL model \cite{forshaw,bartels} and a good
description is obtained. 
\section{Electroproduction Measurements at High ${\bf |\lowercase{t}|}$}

In electroproduction the normalised production and angular 
decay distribution in the $\gamma^{*} p$ centre-of-mass system is a
function of fifteen linear combinations of the spin density matrix elements. 

\begin{figure}
\vspace{7.5cm}
\hspace{-1.0cm}\begin{picture}(14,6)
\put(20.0,235.3){a)}
\put(255.0,235.3){b)}
\put(1.0,-18.3){\epsfig{figure=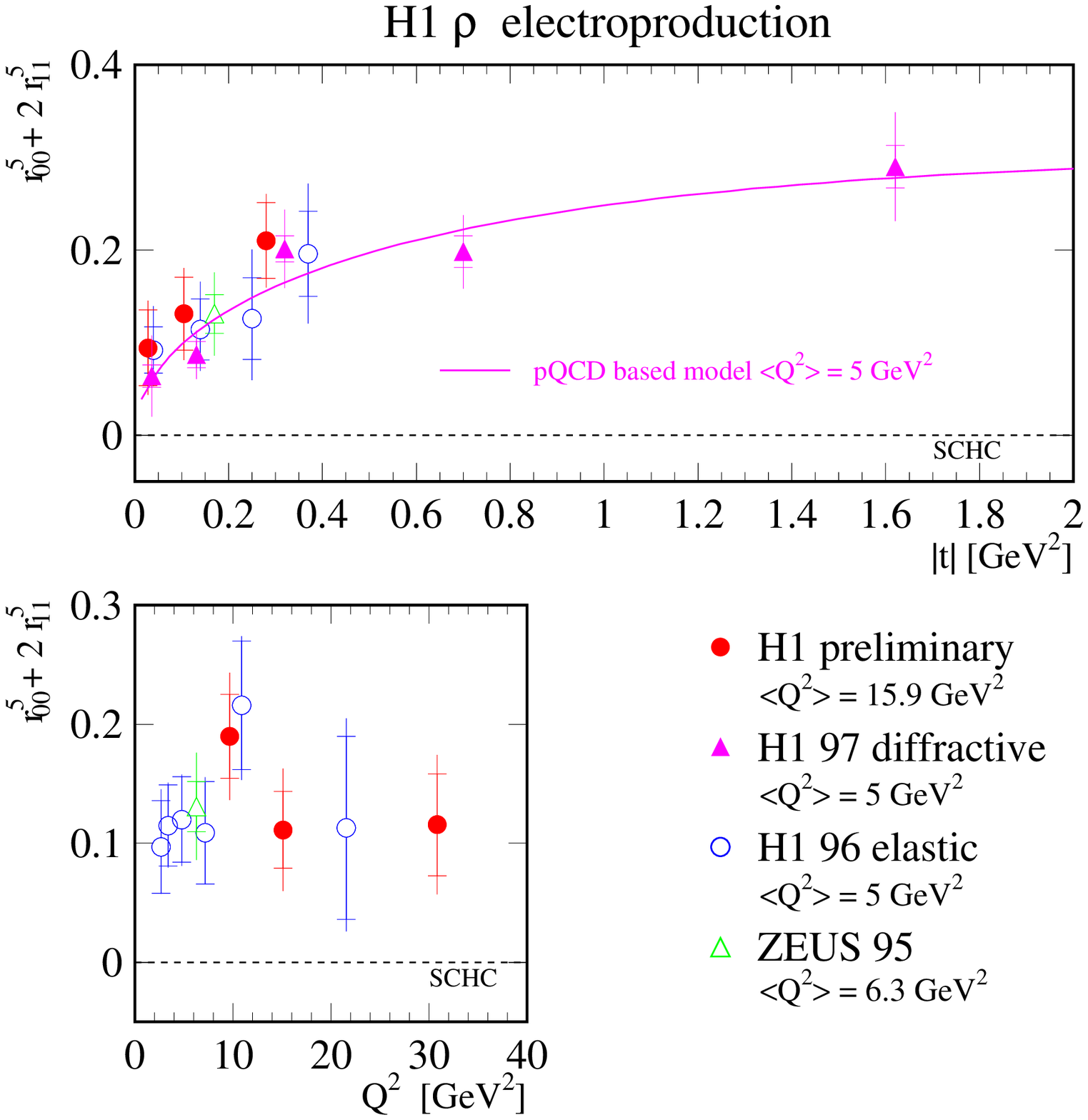,width=9.3cm}}
\put(1,-18.3){\epsfig{figure=whitebox.eps,width=11.3cm,height=4.5cm}}
\put(5.6,-18.3){\hspace{8.1cm}\epsfig{figure=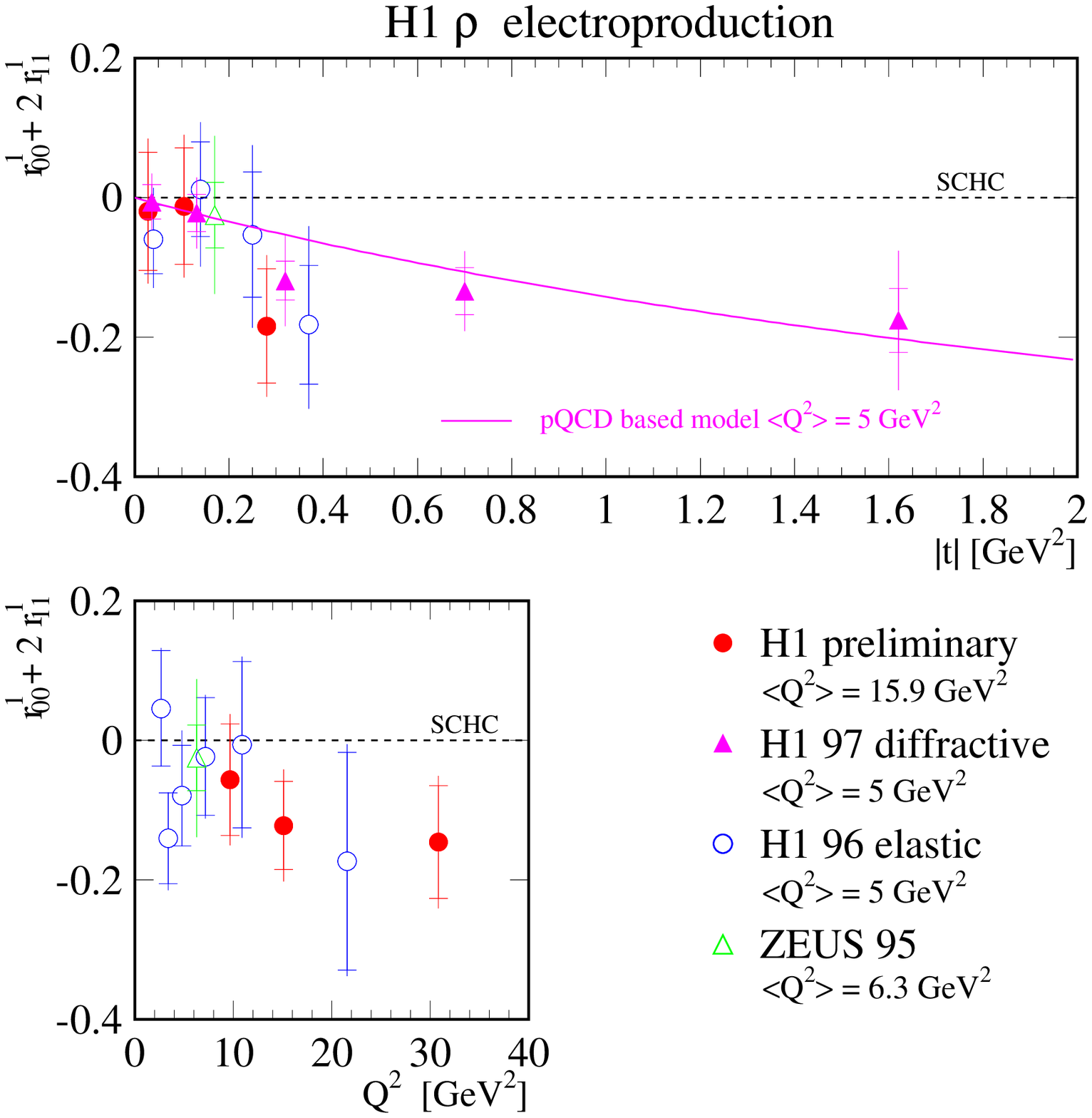,width=9.3cm}}
\put(5.6,-18.3){\hspace{8.1cm}\epsfig{figure=whitebox.eps,width=11.3cm,height=4.5cm}}
\end{picture}
\vspace{-3.8cm}
\caption{High $|t|$ electroproduction $\rho$ measurements of a) $r_{00}^{5}+2r_{11}^{5}$ and b) $r_{00}^{1}+2r_{11}^{1}$ as a function
of $|t|$. The solid line shows the result of a simultaneous fit to the
data (closed triangles) for all
15 linear combinations of the spin density matrix elements.
\label{fig:polar}}
\vskip -0.3cm
\end{figure}

\newpage
The combinations $r_{00}^{5}+2r_{11}^{5}$ and $r_{00}^{1}+2r_{11}^{1}$ 
are presented as a function of $|t|$ in figures \ref{fig:polar}a)
and b) respectively.\cite{Adloff:2002tb} In pQCD the $|t|$ dependence of 
the parameters $r_{00}^{5}$ and $r_{11}^{5}$
are both expected to be proportional to $\sqrt{|t|}$ and the parameters
$r_{00}^{1}$ and $r_{11}^{1}$ proportional to $|t|$ differing
 only in their sign. The result of a simultaneous fit 
to all 15 linear combinations is shown by the solid
line. In figure \ref{fig:polar}a) the violation of SCHC increases 
proportional to $\sqrt{|t|}$ as expected. In figure \ref{fig:polar}b) 
the violation is proportional to $|t|$ as expected. The parameter 
$r_{00}^{1}$ (which depends on the single-flip
amplitude $T_{01}$) dominates due to the negative proportionality.

\section{Summary}
The DVCS cross section exhibits a similar steep rise as a function of
the interaction energy $W_{\gamma p}$ as measured for 
elastic $\rm J/\psi$ photoproduction. The cross section is described by models,
including those incorporating effects of skewed parton
distributions, however when the $Q^{2}$ dependence is
extrapolated assuming a $Q^{2}$ independent $|t|$ dependence these
models lie below the data at large $Q^{2}$. A similar steep 
energy dependence is measured in diffractive $J/\psi$ photoproduction 
at large $|t|$ in two $|t|$
regions. The leading-order BFKL model gives a reasonable simultaneous 
description of the $W_{\gamma p}$ and $|t|$ dependences.  The high $|t|$
dependence of the proton dissociative $\rho$ and $\phi$ production
cross sections are approximately power-like $|t|^{-n}$ with an 
exponent $n \sim 3$ for both the $\rho$ and the $\phi$. Measurements of
the $|t|$ dependences of spin density matrix elements in 
electroproduction $\rho$ measurements agree with pQCD expectations 
showing significant deviations from s-channel helicity conservation.

\end{document}